\documentclass{aastex631}

\usepackage{xcolor}

\usepackage{graphicx}
\usepackage[space]{grffile}
\usepackage{latexsym}
\usepackage{textcomp}
\usepackage{longtable}
\usepackage{multirow,booktabs}
\usepackage{amsfonts,amsmath,amssymb}
\usepackage{url}
\usepackage{hyperref}

\newcommand{\src}{GX~340$+$0}

\newcommand{\nicer}{\textit{NICER}}
\newcommand{\astr}{\textit{AstroSat}}
\newcommand{\ixpe}{\textit{IXPE}}

\newcommand{\hxmt}{\textit{Insight}-HXMT}
\newcommand{\atca}{\textit{ATCA}}
\newcommand{\gmrt}{\textit{GMRT}}

\defcitealias{Bhargava2023ApJ...955..102B}{B23}
\defcitealias{Bhargava2024arXiv240519324B}{Paper I}

\begin{document}

\title{X-ray and Radio Campaign of the Z-source GX 340+0 II: the X-ray polarization in the normal branch}

\author[0000-0002-5967-8399]{Yash Bhargava}
\affiliation{Department of Astronomy and Astrophysics, Tata Institute of Fundamental Research, \\
1 Homi Bhabha Road, Colaba, Mumbai 400005, India}
\email{yash.bhargava\_003@tifr.res.in}

\author[0000-0002-7930-2276]{Thomas D. Russell}
\affiliation{INAF, Istituto di Astrofisica Spaziale e Fisica Cosmica, Via U. La Malfa 153, I-90146 Palermo, Italy}

\author[0000-0002-0940-6563]{Mason Ng}
\affiliation{Department of Physics, McGill University, 3600 rue University, Montréal, QC H3A 2T8, Canada}
\affiliation{Trottier Space Institute, McGill University, 3550 rue University, Montréal, QC H3A 2A7, Canada}

\author[0000-0003-0477-7645]{Arvind Balasubramanian}
\affiliation{Indian Institute of Astrophysics, Koramangala II Block, Bangalore 560034, India}

\author[0000-0003-4498-9925]{Liang Zhang}
\affiliation{Key Laboratory of Particle Astrophysics, Institute of High Energy Physics, Chinese Academy of Sciences, Beijing 100049, China}

\author[0000-0002-2381-4184]{Swati Ravi}
\affiliation{MIT Kavli Institute for Astrophysics and Space Research, Massachusetts Institute of Technology, Cambridge, MA 02139, USA}

\author[0009-0005-0366-0834]{Vishal Jadoliya}
\affiliation{Department of Physics, Indian Institute of Technology Hyderabad, IITH main road, Kandi 502284}

\author[0000-0002-6351-5808]{Sudip Bhattacharyya}
\affiliation{Department of Astronomy and Astrophysics, Tata Institute of Fundamental Research, \\
1 Homi Bhabha Road, Colaba, Mumbai 400005, India}

\author[0000-0002-5900-9785]{Mayukh Pahari}
\affiliation{Department of Physics, Indian Institute of Technology Hyderabad, IITH main road, Kandi 502284}

\author[0000-0001-8371-2713]{Jeroen Homan}
\affiliation{Eureka Scientific, Inc., 2452 Delmer Street, Oakland, CA 94602, USA}

\author[0000-0002-6492-1293]{Herman L. Marshall}
\affiliation{MIT Kavli Institute for Astrophysics and Space Research, Massachusetts Institute of Technology, Cambridge, MA 02139, USA}

\author[0000-0001-8804-8946]{Deepto Chakrabarty}
\affiliation{MIT Kavli Institute for Astrophysics and Space Research, Massachusetts Institute of Technology, Cambridge, MA 02139, USA}

\author[0000-0002-0426-3276]{Francesco Carotenuto}
\affiliation{Astrophysics, Department of Physics, University of Oxford, Keble Road, Oxford OX1 3RH, UK}
\affiliation{INAF-Osservatorio Astronomico di Roma, Via Frascati 33, I-00076, Monte Porzio Catone (RM), Italy}

\author[0009-0005-6441-0468]{Aman Kaushik}
\affiliation{Department of Astronomy and Astrophysics, Tata Institute of Fundamental Research, \\
1 Homi Bhabha Road, Colaba, Mumbai 400005, India}









\begin{abstract}

We present the first X-ray polarization measurement of the neutron star low-mass X-ray binary and Z-source, \src, in the normal branch (NB) using a 200~ks observation with the Imaging X-ray Polarimetric Explorer (\ixpe). This observation was performed in 2024 August. Along with \ixpe, we also conducted simultaneous observations with \nicer, \astr, \hxmt, \atca, and \gmrt\ to investigate the broadband spectral and timing properties in the X-ray and radio wavelengths. During the campaign, the source traced a complete Z-track during the \ixpe\ observation but spent most of the time in the NB.  We measure X-ray polarization degree (PD) of $1.3\pm0.3\%$ in the 2--8 keV energy band with a polarization angle (PA) of $38\pm6^\circ$. The PD in the NB is observed to be weaker than in the horizontal branch (HB) but aligned in the same direction. 
The PD of the source exhibits a marginal increase with energy while the PA shows no energy dependence. The joint spectro-polarimetric modeling is consistent with the observed X-ray polarization originating from a single spectral component from the blackbody or the Comptonized emission, while the disk emission does not contribute towards the X-ray polarization. \gmrt\ observations at 1.26~GHz during HB had a tentative detection at 4.5$\pm$0.7 mJy while \atca\ observations a day later during the NB detected the source at 0.70$\pm$0.05 mJy and 0.59$\pm$0.05 mJy in the 5.5 \& 9 GHz bands, respectively, suggesting an evolving jet structure depending on the Z-track position.

\end{abstract}



\section{Introduction} \label{sec:intro}


Z-sources are a unique subclass of accreting neutron star (NS) low mass X-ray binaries (LMXBs). These sources trace a characteristic Z-shaped track on their X-ray Hardness-Intensity diagrams (HIDs) or color-color diagrams (CCDs) and based on their brightness and estimated distances, are expected to accrete close to the Eddington limit \citep{vdkReview2004astro.ph.10551V}. These sources host a low magnetic field NS ($10^{8-9}$~G). 
The spectrum of NS XRBs have been historically modeled as either a combination of a multicolor blackbody (e.g., an accretion disk) and a non-thermal component \citep[Eastern model, e.g.,][]{mitsuda1989PASJ...41...97M} or a single temperature blackbody (surface emission from the NS) and a non-thermal component \citep[Western model, e.g.,][]{White1988ApJ...324..363W}. 
The non-thermal component in either case is often modeled as a Comptonized plasma of hot electrons and could have either a shell-like geometry (where the hot plasma covers the NS and up-scatters the photons from the surface; e.g., \citealt{gnarini2022MNRAS.514.2561G}) or a slab-like geometry (where the plasma is like a sandwich around the accretion disk; e.g., \citealt{gnarini2022MNRAS.514.2561G}). The emission could also arise from the boundary layer (BL), due to matter between the accretion disk and the NS \citep[][]{Shakura1988AdSpR...8b.135S}, or the spreading layer (SL) where matter accreted from the disk partially covers the NS surface \citep{SLlapidus1985MNRAS.217..291L, SLinogamov1999AstL...25..269I}. 

The characteristic Z-track can be divided into three main branches: the horizontal branch (HB), the normal branch (NB) and the flaring branch (FB) with the hard apex connecting HB and NB and the soft apex connecting  NB and  FB \citep{hasinger1989A&A...225...79H}. Some Z-sources also show a branch beyond the FB, often called the extended flaring branch \citep[EFB;][]{Jonker2000ApJ...537..374J,Church2010MmSAI..81..275C, Gibiec2011xru..conf..212G}. In the known set of NS LMXBs, only a few sources persistently show Z-source like behavior: Sco~X-1, GX~17$+$2, GX~349$+$2, Cyg~X-2, GX~5$-$1, \src, GX~13$+$1 and LMC~X-2 \citep{hasinger1989A&A...225...79H, Smale2003ApJ...590.1035S, fridriksson2015, Kaddouh2024RNAAS...8..243K} while some sources (XTE~J1701$-$462, IGR~J17480$-$2446 and Cir X-1) have intermittently shown a Z-track in their HID indicating that these tracks are strongly related to changes in the accretion rate \citep{homan2007ApJ...656..420H, lin2009ApJ...696.1257L, Homan2010ApJ...719..201H, Chakrabortyetal2011, fridriksson2015,2024ApJ...966..232N}. The shapes of the `Z' on the HID are canonically of two types: Cyg-like, where the track length of the HB and the NB are roughly similar \citep[seen in Cyg X-2, GX 5-1 and \src;][]{hasinger1989A&A...225...79H} and Sco-like where the track lengths of the NB and the FB are similar and the HB is significantly shorter \citep[seen in Sco X-1, GX 17+2, GX 349+2, GX 13+1 and LMC X-2;][]{hasinger1989A&A...225...79H,Smale2003ApJ...590.1035S,Kaddouh2024RNAAS...8..243K}.


The jets observed from Z-sources show strong variability as the mass accretion rate changes as it traverses along its Z-shaped path in the HID  \citep[e.g.,][]{1986ApJ...306L..91P,1990ApJ...365..681H,1990A&A...235..147H}. This variability arises as the structure and properties of the jet change with the changing mass accretion rate, like the state transitions detected in other XRBs \citep[e.g.,][]{2004MNRAS.355.1105F,2006MNRAS.366...79M}. In the few cases that dedicated, high-cadence radio monitoring has been carried out on Z-sources, the radio emission from the steady jet is generally seen to increase in luminosity as the source moves along the HB towards the NB \citep[e.g.,][]{2007ApJ...671..706M}. Around the transition from the HB to the NB, bright, flaring, optically-thin radio emission is detected, consistent with the launching of the transient jet. That emission fades throughout the NB, becoming radio faint during the FB. As the source then transitions back through the FB towards the  HB via the NB, the source re-brightens in the radio once again \citep{2001ApJ...553L..27F,2001ApJ...558..283F,2003ApJ...592..486B,2010A&A...512A...9B,2012A&A...546A..35C}. In addition, it has been proposed that ultra-relativistic ejecta may be launched close to the transition between the NB and the FB that cause bright radio flares as the ultra-relativistic ejecta collide with the downstream jet \citep{2004Natur.427..222F,2019MNRAS.483.3686M}.

\src\ is a bright Cyg-like Z-source that shows all Z-track branches. It has demonstrated a quick tracing of the complete Z-track, on a timescale of a few days \citep{Jonker2000ApJ...537..374J}, indicating a rapid change in spectral states. On the other hand, during some of few day-long observations, the source remained stable in one of the branches \citep[e.g. mainly within the NB with rapid excursions to the FB, or mainly in the HB with no/few excursions to the NB;][]{Jonker2000ApJ...537..374J, Seifina2013ApJ...766...63S}. \src\ lies in a direction of high line of sight absorption column \citep[$2.2\times10^{22}$~cm$^{-2}$;][]{nh2016A&A...594A.116H} and the source also has a high intrinsic absorption column \citep[as indicated by $n_{\rm H}\approx(6-12)\times10^{22}$~cm$^{-2}$ in the literature;][ hereafter \citetalias{Bhargava2023ApJ...955..102B}]{Church2006A&A...460..233C, Iaria2006ChJAS...6a.257I, Bhargava2023ApJ...955..102B}. The X-ray spectral modeling revealed presence of multiple emission components, often characterized as a combination of thermal and non-thermal components (\citealt{Church2006A&A...460..233C}, \citealt{Seifina2013ApJ...766...63S}, \citetalias{Bhargava2023ApJ...955..102B}) across all branches. Infrared observations of the source have been able to identify a possible counterpart \citep{Miller1993AJ....106...28M}, while radio observations have shown a strong variation of the radio flux as a function of the Z track position \citep{Penninx1993A&A...267...92P, oosterbroek1994A&A...281..803O, 2000MNRAS.318..599B}. Recent radio observations taken in the HB also indicated a spectral break somewhere between 1.2--5.5 GHz, likely a result of synchrotron self-absorption, or free-free absorption, as well as an evolving radio jet \citep[hereafter \citetalias{Bhargava2024arXiv240519324B}]{Bhargava2024arXiv240519324B}.
\src\ was observed with \ixpe\ in 2024 March when the source was mainly within the HB \citep[\citetalias{Bhargava2024arXiv240519324B},][]{LaMonaca2024arXiv241000972L}. The model independent polarization in the 2--8~keV energy band was estimated to be $4.0\pm0.4$\% with a polarization angle (PA) of $38\pm3^\circ$. The PA of the lowest energy bin (2--2.5~keV) was hinted to be different from rest of the energy band indicating a different origin of the polarized emission. The spectro-polarimetric analysis of joint \ixpe-\astr\ observation using the spectral model from \citetalias{Bhargava2023ApJ...955..102B} suggested that the polarized 2--2.5~keV emission could arise from the accretion disk while the polarized emission above 2.5~keV could be attributed solely to the Comptonized emission or from a contribution of both blackbody and Comptonized emission \citepalias{Bhargava2024arXiv240519324B}. Alternatively, the spectrum of the source can also be decomposed with the Eastern model, where the accretion disk is expected to account for the low energy polarization and the corona contributes to the high energy polarization \citep{LaMonaca2024arXiv241000972L}. 
The observed polarization degree (PD) of \src\ in the HB matches well with the X-ray polarization seen in the HB from other Z-sources \citep[e.g., XTE~J1701$-$462 and GX~5$-$1][respectively]{Cocchi2023A&A...674L..10C,Fabiani2024A&A...684A.137F}. Other, similar, sources that have been observed in the NB and the FB report a lower PD of 0.6--2\% \citep[Cyg X-2, XTE~J1701$-$462, GX~5$-$1, Sco~X-1 and Cir~X-1;][respectively]{Farinelli2023MNRAS.519.3681F, Cocchi2023A&A...674L..10C, Fabiani2024A&A...684A.137F, monaca2024ApJ...960L..11L, rankin2024ApJ...961L...8R} suggesting a strong geometrical change as the spectral states change.      

The \ixpe\ observation of the \src\ reported by \citetalias{Bhargava2024arXiv240519324B} was disrupted due to a spacecraft anomaly, with the rest of the observation occurring from 2024 August 12-16. We supplemented this second \ixpe\ observation with an extensive campaign involving various X-ray and radio facilities.  In this article, we report the results from this second campaign and compare it with the insights from the previous campaign taken in 2024 March. We describe the data reduction procedures for individual instruments in Section~\ref{sec:data_red}, detail the analysis methodologies and results in Section~\ref{sec:data_res}, and interpret the results in Section~\ref{sec:disc}.

\section{Observations and Data reduction} \label{sec:data_red}

\subsection{IXPE}\label{ssec:ixpe}
\begin{figure}
  \includegraphics[width=\columnwidth]{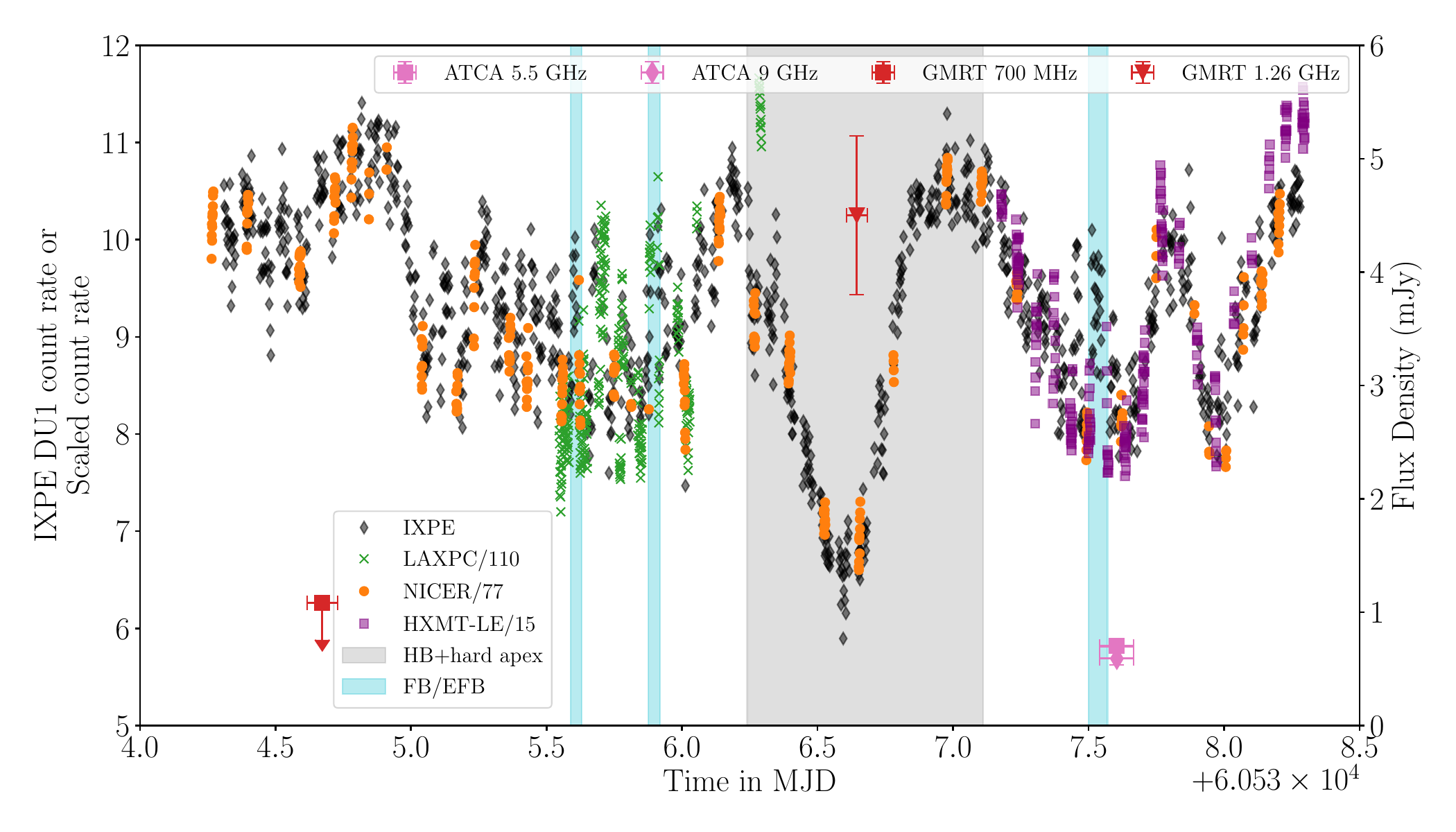}
  \caption{The evolution of \src\ as seen by \ixpe\ is shown as black diamonds. The \ixpe\ observation was supplemented with joint observations with \nicer\ (in orange circles), \astr\ (LAXPC light curve as green crosses), and \hxmt\ (LE light curve as purple squares). The observation log of these instruments is noted in table~\ref{tab:obslog}. The light curves from \nicer, \astr-LAXPC and \hxmt/LE are scaled to highlight the similarity in the joint observations and the scaling factors are mentioned in the legend of the plot. Additionally, we also show results from radio observations conducted during the X-ray campaign and shown as the magenta and red points (where the radio flux density can be read on the right y-axis). The intervals for the FB/EFB and the HB (including the hard apex) are indicated by the cyan and grey shaded regions, respectively. The source was in the NB at all other times (see table~\ref{tab:NB_intervals} for exact time stamps of NB intervals). }\label{fig:lc}
\end{figure}

\ixpe\ conducted a $\approx$200~ks observation of \src\ from 2024 August 12--16. The details of the observations are listed in Table~\ref{tab:obslog}. 
We used \texttt{ixpeobssim} software (v31.0.1) to bin the event data \citep{2015APh....68...45K,ixpeobssim2022SoftX..1901194B} into the high-level products (e.g., energy-dependent light curves, Stokes I, Q, and U spectra, and model-independent polarization properties with the \textsc{pcube} algorithm). The event data were calibrated and weighted according to the calibration files corresponding to recent calibration epoch of 2024 July 1  \citep[\texttt{obssim20240701\_alpha075};][]{dimarco2022AJ....164..103D}  for the lightcurves and spectra and the unweighted calibration files were used for \textsc{pcube} \citep[\texttt{obssim20240701};][]{dimarco2022AJ....164..103D, Forsblom2024A&A...691A.216F}. 
For the extraction of the source photons, we adopted a 1.6$\arcmin$ circular region. Since the count rate of the source was higher than 2\,c/s/arcmin$^2$ (see Figure~\ref{fig:lc}), we do not perform any background rejection or subtraction \citep[as suggested in][]{dimarco2023AJ....165..143D}. For the model-independent analysis, the polarization information from individual DUs were combined and reported in Section~\ref{sssec:pcube}. For the spectral analysis, we re-binned the I spectrum to ensure at least 25 counts per bin (to enable $\chi^2$ statistics).  The Stokes Q/U spectra were not rebinned as they had similar number of bins as the Stokes I spectra in the operational energy range of 2--8~keV. 


\begin{table*}
  \centering
  \caption{Observation log for the multi-wavelength campaign of \src}\label{tab:obslog}
  \begin{tabular}{|ll|c|cc|}
    \hline
    Observatory & Instrument & Observation ID & Start Time & Stop Time  \\ \hline
    \ixpe & GPD (DU1--3) & 03009901 & 2024-08-12 07:26:35 & 2024-08-16 06:45:03 \\ \hline
    \multirow{5}{*}{\nicer} & \multirow{5}{*}{XTI} & 7705010101 & 2024-08-12 06:20:47 & 2024-08-12 21:51:37  \\
     &  & 7705010102 & 2024-08-13 00:55:35 & 2024-08-13 21:03:55 \\
     &  & 7705010103 & 2024-08-14 00:08:50 & 2024-08-14 23:31:20 \\
     &  & 7705010104 & 2024-08-15 02:27:38 & 2024-08-15 22:41:48 \\
     &  & 7705010105 & 2024-08-16 00:08:37 & 2024-08-16 04:56:07 \\ \hline
    \astr& LAXPC & 9000006386 & 2024-08-13 13:07:31 & 2024-08-14 07:01:41   \\ \hline
    \multirow{2}{*}{\hxmt} & LE &P061437700402- & 2024-08-15 04:14:58 & 2024-08-16 07:07:50  \\
      & ME  & P061437700410& 2024-08-15 04:14:28 & 2024-08-16 07:57:19 \\ \hline
     
    \gmrt & Band 4 & ddtC367 & 2024-08-12 14:47:47 & 2024-08-12 17:29:42 \\ 
    \gmrt  & Band 5 & ddtC367 & 2024-08-14 14:32:23 & 2024-08-14 16:24:42 \\ 
     \atca & -- &  & 2024-08-15 13:43:40 & 2024-08-15 15:49:40 \\ \hline
     
  \end{tabular}
\end{table*}

\subsection{NICER}\label{ssec:nicer}

The Neutron star Interior Composition Explorer \citep[\nicer;][]{Gendreau2016SPIE.9905E..1HG} is a non-imaging instrument onboard the International Space Station, comprising 56 silicon drift detector and X-ray concentrator optic pairs (52 operational) in focal plane modules (FPM). \nicer\ observed \src\ between 2024 August 12 and 2024 August 16 (see Table~\ref{tab:obslog}), within the IXPE observation interval. The raw data were reduced and processed using \texttt{HEASOFT} version 6.33 and the \nicer\ Data Analysis Software (\texttt{NICERDAS}) version 12 (2024-02-09\_V012) using calibration version \texttt{xti20240206}. We employed the following filtering criteria in generating the good time intervals: \nicer\ being outside of the South Atlantic Anomaly; angular offset for the source of $<54\arcsec$; Earth limb elevation angle ELV $>20\degr$; undershoot rate of 0--500 c/s/FPM; overshoot rate of 0-30 c/s/FPM, and a minimum cutoff rigidity such that \texttt{COR\_SAX} $>$ 1.5. The processing was done using the \texttt{threshfilter=DAY}  as the observations were taken during the orbital day and thus the lowest energy data may be affected.
To investigate the branch-resolved behavior of the source, we first combined the observations using \texttt{niobsmerge}. 
These filtering criteria resulted in a total exposure of 15.3~ks for scientific analysis. We generated spectral products (including response matrices) using \texttt{nicerl3-spect}, where we grouped the spectra with the optimal binning scheme \citep{2016A&A...587A.151K} and rebinned the spectra such that each bin had a minimum of 25 counts. We also generated light curves in various energy bands (the 0.5--10 keV light curve binned at 50~s is shown in Figure~\ref{fig:lc} in orange circles) using \texttt{nicerl3-lc}.
Using the time intervals for which we identified the source to be in the different branches (see Section \ref{ssec:state_iden} for details on the state identification and Table~\ref{tab:NB_intervals} for the NB intervals),
we constructed the spectrum from the merged event file for the NB intervals and extracted the appropriate response and auxiliary response files. 
For estimating the background, we made use of the SCORPEON model \citep{SCORPEON2024HEAD...2110536M}. 

\subsection{AstroSat}\label{ssec:astrosat}
To supplement the \ixpe\ observation with broad X-ray energy coverage, we conducted a ToO observation with \astr\ \citep{Singh2014SPIE.9144E..1SS}, running 2024 August 13--14. For the observation we used the Large Area X-ray Proportional Counter \citep[LAXPC;][]{Yadav2016SPIE.9905E..1DY,Yadav2017CSci..113..591Y} as the primary instrument.
We procured the level 1 data of the LAXPC observation from AstroBrowse\footnote{\url{https://astrobrowse.issdc.gov.in/astro_archive/archive/Home.jsp}} and reduced the data using relevant tools in \textsc{LAXPCsoftware22Aug15}\footnote{\url{http://astrosat-ssc.iucaa.in/uploads/laxpc/LAXPCsoftware22Aug15.zip}} \citep{Antia2021JApA...42...32A,Misra2021JApA...42...55M}. 
The pipeline also includes the tools to filter the South Atlantic Anomaly passage and Earth occultation intervals and extract the energy-dependent light curves, spectra, background spectra, and background light curves. For the analysis, we considered only LAXPC 20 as LAXPC 30 was turned off early in the mission and LAXPC 10 indicated abnormal gain variation. To minimize the background contribution over 3-20~keV, we used the layer 1 data for the extraction of higher-level products (e.g., light curves and spectra). We show the scaled light curve (binned at 50~s) from the LAXPC observation in Figure~\ref{fig:lc} as green crosses. The spectrum from LAXPC was rebinned according to the detector response. Due to response uncertainties, we included a 2\% systematic error in the spectrum across the complete energy range (3--20~keV). 


\subsection{Insight-HXMT}\label{ssec:hxmt}

\hxmt\ \citep{Zhang2020SCPMA..6349502Z} observed \src\ from 2024 August 15, 00:44:17 to 2024 August 16, 08:28:41. The data were extracted from all three instruments using the \hxmt\ Data Analysis software (HXMTDAS) v2.06\footnote{The data analysis software is available from \url{http://hxmten.ihep.ac.cn/software.jhtml}.}, 
and filtered with the following standard criteria: (1) pointing offset angle less than $0.04^{\circ}$; (2) Earth elevation angle larger than $10^{\circ}$; (3) the value of the geomagnetic cutoff rigidity larger than 8 GV; (4) at least 300 s before and after the South Atlantic Anomaly passage. To avoid possible contamination from the bright Earth and nearby sources, we only use data from the small field of view (FoV) detectors \citep{Chen2018ApJ...864L..30C}. The software also provides the relevant response and background files for the spectral analysis.  The HE data are not used in our spectral analysis as the HE spectra were dominated by the background. The energy bands adopted for spectral analysis are 2--10~keV for LE and 10--20~keV for ME. The scaled light curves from \hxmt-LE in 3--10~keV and binned at 50~s are shown in figure~\ref{fig:lc} as purple circles.

\subsection{GMRT}\label{ssec:gmrt}
We observed \src\ with the Giant Metrewave Radio Telescope (\gmrt). Utilizing the wideband receiver backend of \gmrt, data were recorded in two frequency bands: band 4 (central frequency 750~MHz, bandwidth 400~MHz) and band 5 (central frequency 1260~MHz, bandwidth 400~MHz) on 2024 August 12 and 2024 August 14, respectively (Proposal ddtC367, for simultaneous coverage with the \ixpe\ campaign). 
The raw data were downloaded in the \texttt{FITS} format and converted to the Common Astronomy Software Applications \citep[\texttt{CASA};][]{CASA2022PASP..134k4501C} measurement set format. The data were calibrated (3C286 was used for flux and bandpass calibration, and J1717-398 was used for phase calibration) and imaged using the automated continuum imaging pipeline \texttt{CASA-CAPTURE} \citep{Kale2021ExA....51...95K}. Manual flagging and calibration were performed as required. Eight rounds of self-calibration (both phase and amplitudes) were completed within each pipeline run to obtain the final cleaned images. No sources were detected at the position of \src\ in the band 4 observations, providing a 3$\sigma$ upper limit of 1.1~mJy at 750~MHz, where we use the root mean square (RMS) of the noise from a large area in a source-less region of the image (close to the source position) as $\sigma$. The band 5 data were heavily affected by radio frequency interference (RFI), but the image shows a tentative point source close to the known position of \src\ with a flux density, $S_{\nu}$, of 4.5$\pm$0.7\,mJy at 1.26\,GHz (10\% flux density error added in quadrature to the RMS in a large source-less region to obtain the error in the maximum).

\subsection{ATCA}\label{ssec:atca}
The Australia Telescope Compact Array (ATCA) observed \src\ on 2024 August 15 between 13:43:40 and 15:49:40 UT. During the observation ATCA was in a relatively extended 1.5~km configuration\footnote{\url{https://www.narrabri.atnf.csiro.au/operations/array_configurations/configurations.html}}. 
Data were recorded simultaneously at central frequencies of 5.5 and 9\,GHz, with 2\,GHz of bandwidth at each frequency band. We used PKS~B1934$-$638 for flux and bandpass calibration, and the nearby calibrator J1631$-$4345 for phase calibration. Data were flagged, calibrated, and imaged using standard procedures within the (\texttt{CASA} version 5.3.1; \citealt{CASA2022PASP..134k4501C}). Imaging used a Briggs robust parameter of 0, balancing sensitivity and resolution. Fitting for a point source in the image plane, \src\ was detected at both central frequencies, where $S_{\nu} = 0.70 \pm 0.05$\,mJy at 5.5\,GHz and $0.59 \pm 0.05$\,mJy at 9\,GHz. This corresponds to a radio spectral index, $\alpha$, of $-0.35 \pm 0.25$ (defined as $S_{\nu} \propto \nu^{\alpha}$, where $\nu$ is the frequency). Such a radio spectral index could be consistent with a steep spectrum from transient ejecta, or a flat radio spectrum from a steady jet. 

To explore intra-observational variability, we also imaged the source on 20-minute timescales. We found that at both central frequencies the flux density decreased for the first $\sim$half of the observation, before remaining stable during the second half. At 5.5\,GHz the flux density decreased from $1.07 \pm 0.15$\,mJy at the start of the observation to $0.44 \pm 0.10$\,mJy by the middle of the observation. During the second half of the observation, the 5.5\,GHz flux density then remained stable (within errors), measuring $0.50 \pm 0.10$\,mJy at the end of the observation. At 9\,GHz, the flux density was measured to be $0.82 \pm 0.08$\,mJy at the start, dropping to $0.32 \pm 0.06$\,mJy by the middle of the observation, remaining stable (within errors) to the end of the observation $0.34 \pm 0.10$\,mJy.


\section{Data analysis and Results} \label{sec:data_res}

\subsection{Identification of the states}\label{ssec:state_iden}
\begin{figure*}
  \centering
  \includegraphics[width=0.45\columnwidth]{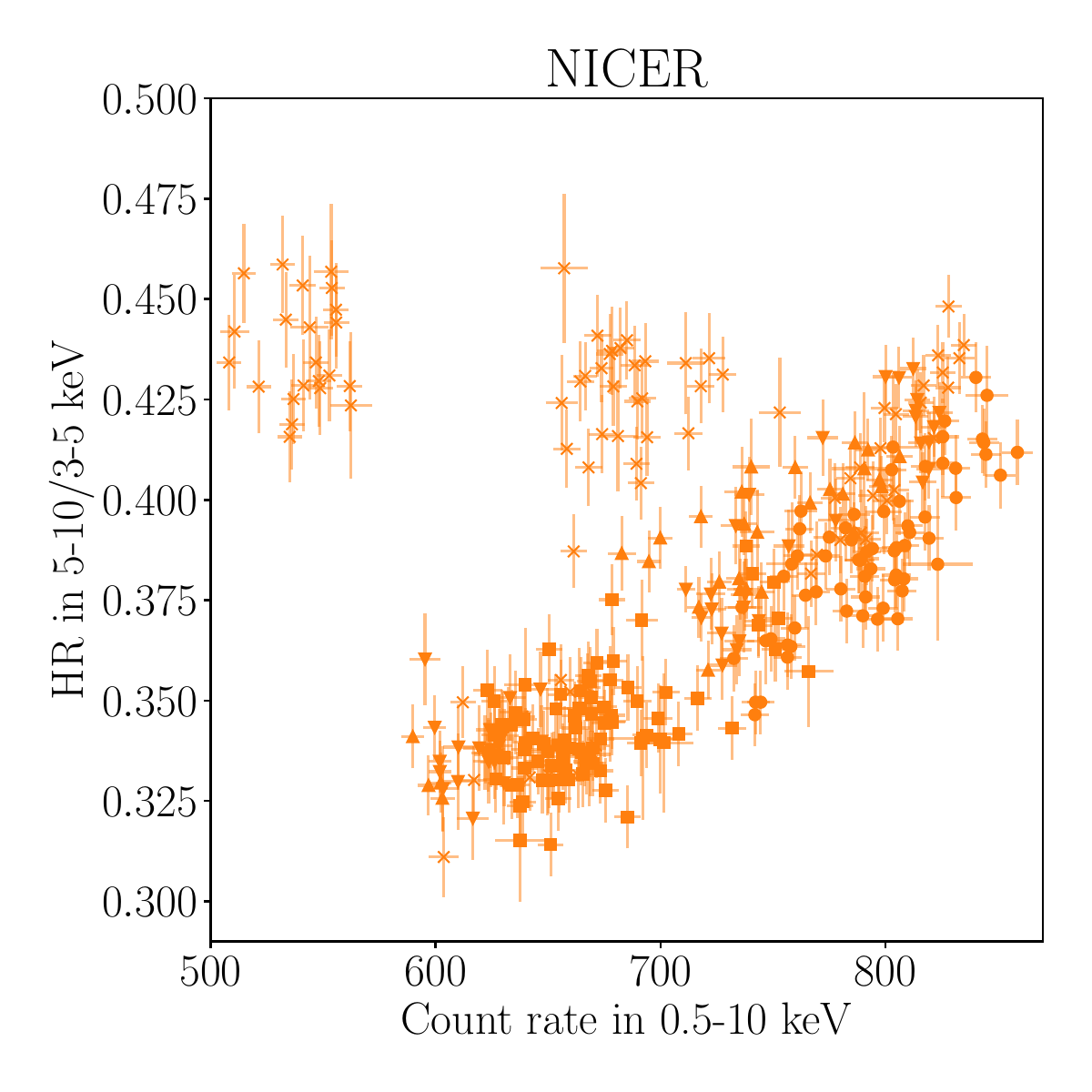} \includegraphics[width=0.45\columnwidth]{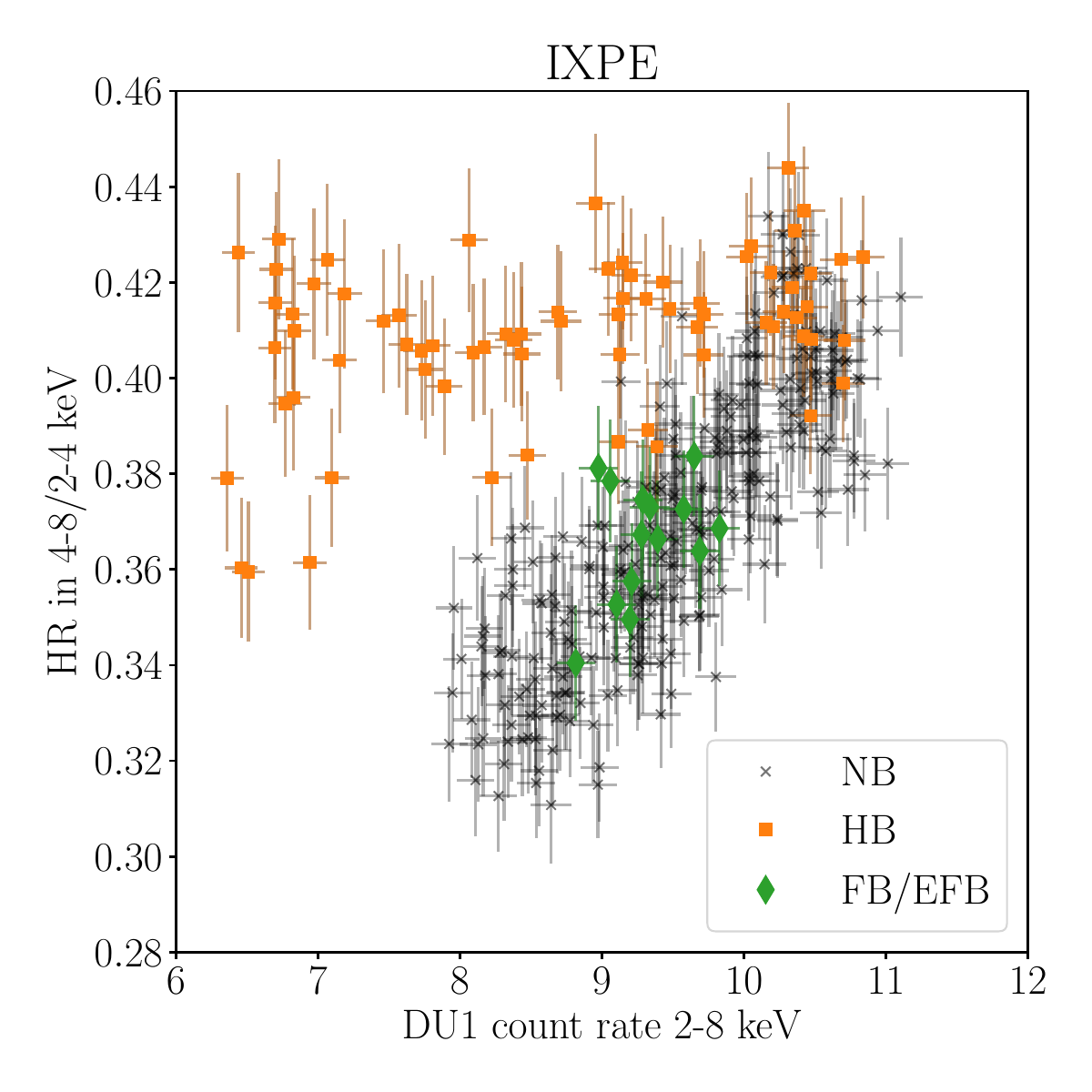} 
  \includegraphics[width=0.45\columnwidth]{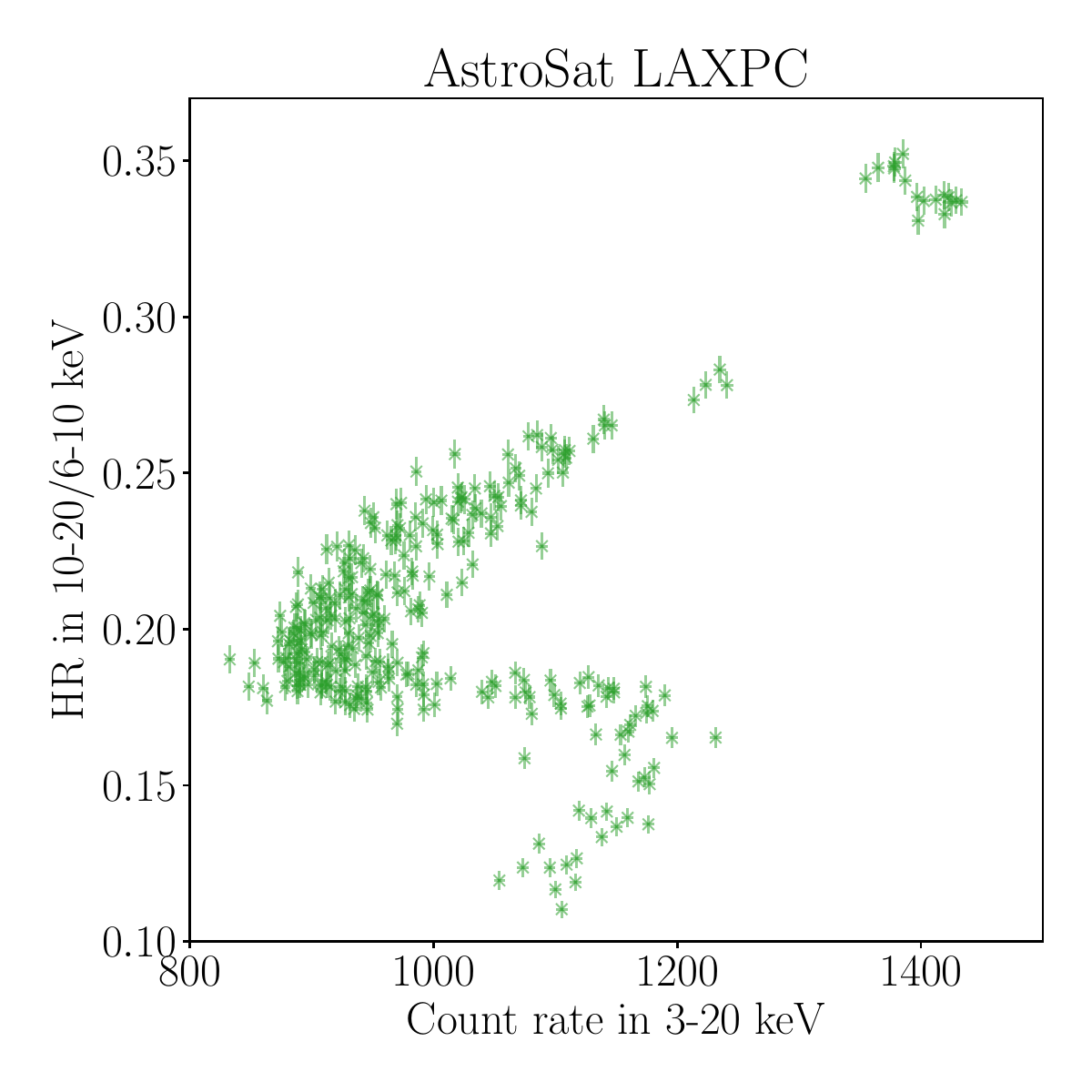}\includegraphics[width=0.45\columnwidth]{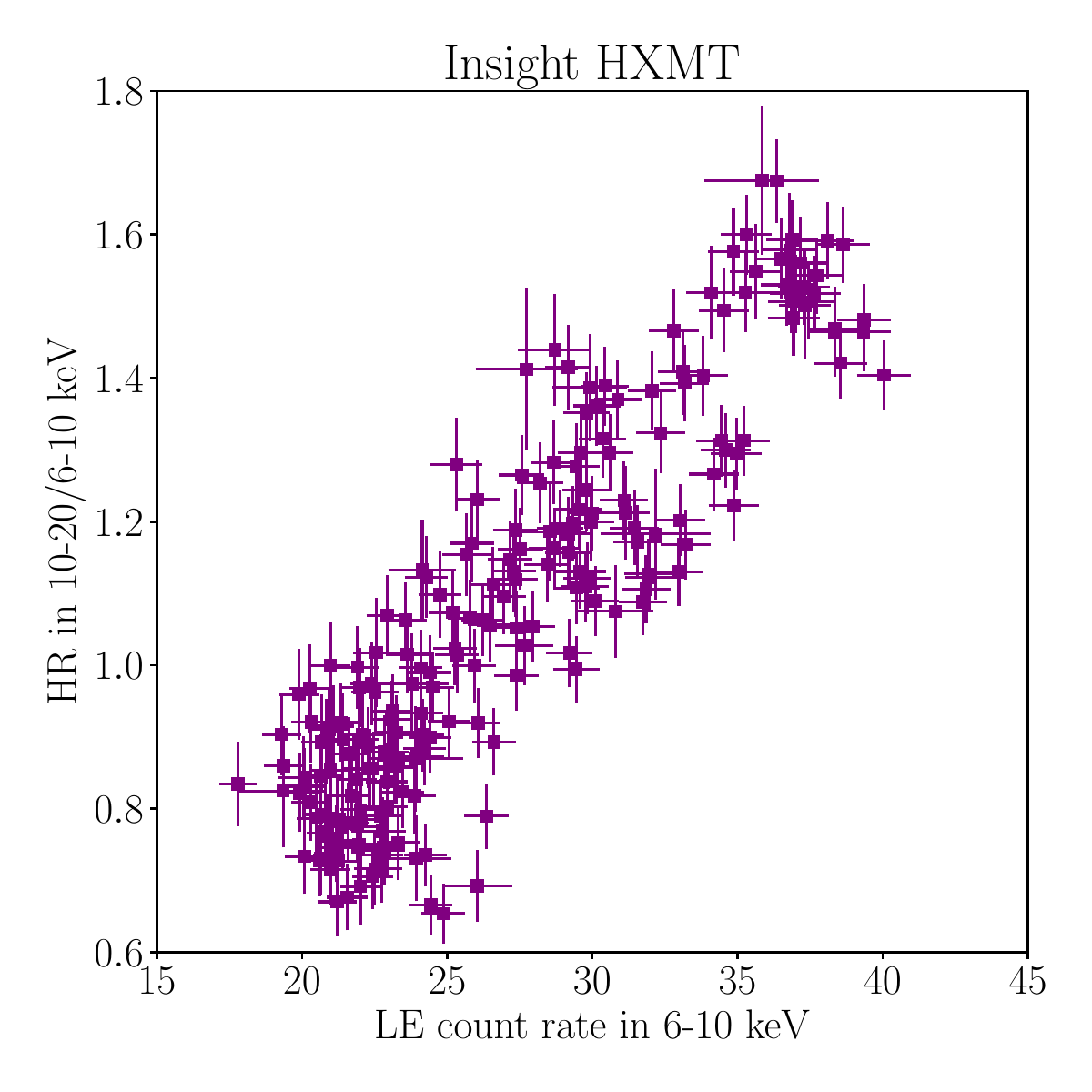} 
  \caption{Hardness intensity diagrams (HIDs) of \src\ as seen by different observatories during the X-ray and radio campaign. The \nicer\ and \astr\ HIDs (top left and bottom left, respectively) indicate that the source traversed through all the states during the campaign. Comparing the HID and light curve in Figure~\ref{fig:lc}, we can infer that the source went into the HB only once (during the big dip from MJD 60536.24-60537.11). During the rest of the time, the source was in the NB with a few excursions into the FB and EFB. The HID from the \ixpe\ observation (top right) is able to distinguish the HB (shown as the orange  squares) from the NB (black crosses) around the apex.  In addition, the \ixpe\ points that lie within the FB/EFB intervals, as identified from \astr\ and \hxmt, are shown with green diamonds and clearly not identifiable solely from the \ixpe\ HID.
  The \hxmt\ HID is generated by taking a ratio of ME in 10--20~keV and LE in 6--10~keV to see if the source underwent a transition to FB. The HID seems to indicate a slight bend into FB. }\label{fig:hid}
\end{figure*}

Previous IXPE observations of Z-sources have indicated that the X-ray polarization from these systems is dependent on the branch position. Thus, we need to identify the source state throughout the observation to interpret the nature of X-ray polarization and therefore the emission components. We constructed the HID using all the X-ray instruments using different definitions of hardness (typically limited by the energy range from the instruments) and have shown them in Figure~\ref{fig:hid}. 

From the HID, \src\ traced the complete Z-track during the observation. The HID from \nicer\ (count rate in 3--10~keV and HR as the ratio between 5--10 and 3--5~keV bands) can identify the epochs when the source was in the HB and the NB while the HID from \astr-LAXPC (count rate in 3--20~keV and HR as the ratio between 10--20 and 6--10~keV bands) highlights the track from the hard apex to the EFB. The \hxmt\ observation (HID constructed using the LE and ME light curves in the 6--10 and 10--20~keV bands, respectively) were mainly in the NB but for some duration the source seemed to indicate an excursion in the FB (as seen below the HR of 0.8 in bottom right panel of figure~\ref{fig:hid}).  The identified FB/EFB and HB intervals are marked in the \ixpe\ HID. Notably, the FB/EFB intervals are not distinguishable from NB solely based on the HID (see appendix~\ref{appsec:soft_HID} for a deeper investigation). 

For this study, we identified FB intervals using the \astr-LAXPC and \hxmt\ HID and extrapolate the intervals to the data gaps (if present) using the fact that Z-sources continuously move along the Z-track without jumping between branches \citep[e.g.][\citetalias{Bhargava2023ApJ...955..102B}]{Church2006A&A...460..233C,homan2007ApJ...656..420H, sriram2011ApJ...743L..31S, lin2012ApJ...756...34L}. Since we focus on analysis of the NB, these intervals are excluded from any estimation of the polarization or the spectrum.  A possible caveat in the state identification is that during the intervals where the source was covered by only \nicer\ and \ixpe, there may be some FB/EFB intervals which have been misclassified as NB. The similarity of the FB/EFB and NB in the HID in the soft X-ray emission (e.g., Figure~\ref{fig:hid_astrosat_soft}) would suggest a similarity in the spectrum of these two branches (at least $\lesssim$10~keV). Since the states seem to be indistinguishable in the HID, we assume that the inclusion of these segments during the NB doesn't affect the spectral or polarimetric results significantly.

\begin{table}
  \centering
  \caption{The start and end times of identified NB intervals. The intervals mentioned here include the data gaps evident in the light curves (see Figure~\ref{fig:lc})} \label{tab:NB_intervals}
  \begin{tabular}{|cc|}
  \hline
    Start MJD & End MJD \\ \hline
    60534.30  &  60535.59 \\
    60535.63 &  60535.69\\
    60535.72 & 60535.87\\
    60535.92 &  60536.24 \\
    60537.11 &  60537.50  \\
    60537.57 &  60538.40 \\ \hline
  \end{tabular}
\end{table}


\subsection{Spectral modeling of NB}\label{ssec:spec_mod}

To understand various spectral components from the \src\ during the NB, we extracted the spectra for all instruments using the methods described in their respective sections (Sections \ref{ssec:ixpe}-\ref{ssec:hxmt}) for the intervals corresponding to the NB (Table~\ref{tab:NB_intervals}). During the epochs of the \astr\ and \hxmt\ observations, we excluded any FB/EFB intervals as determined in section \ref{ssec:state_iden}. We  conducted the spectral and spectro-polarimetric modeling in \textsc{xspec v12.14.1}. For fitting  of the spectra, we used \texttt{PGstat} for \nicer\ spectra (as the background estimation for \nicer\ observations is done by SCORPEON modeling and results in non-Poissonian background) and $\chi^2$ for rest of the spectra. We used $\chi^2$ as the test statistic and report $1\sigma$ confidence intervals for the estimated parameters. 

Since the focus of the analysis is to identify the spectral components responsible for the X-ray polarization, we utilized the joint coverage of \ixpe\ and \nicer\ to identify the spectral components. We also included the spectrum from \astr-LAXPC as it overlapped with \nicer\ and \ixpe\ and allows us to extend the spectral range to 20~keV, placing a strong constraint on the hard X-ray emission, primarily the non-thermal component. 
At the softest energies ($<$0.6~keV), the \nicer\ spectrum was dominated by the noise peak due to additional optical loading from the orbital day. Thus we ignored the \nicer\ spectrum below 0.6~keV. In case of spectra from other instruments, we excluded the energy intervals where the background counts were higher than the source counts. We note that inclusion of \hxmt\ spectra in the joint fit resulted in bad fit for all instruments. A better fit was obtained if we allowed one set of parameters for \ixpe, \nicer, and \astr-LAXPC and one set of parameters for \hxmt, which could be due to calibration mismatch between these instruments. 
Since the inclusion of \hxmt\ spectra did not provide any additional information over joint \ixpe, \nicer, and \astr-LAXPC spectra, we report the analysis of the latter set of spectra.  The individual \ixpe-DU spectra suggested differences in the calibration and therefore we left the gain slope and offset parameter free during the initial modeling \citep[e.g.][]{LaMonaca2024arXiv241000972L} and fixed it to best-fit values for computation of the confidence intervals on the spectral parameters. The typical gain offset for each spectra was found to be $\lesssim5\%$ (see Table~\ref{tab:spec_pars} for the estimated gain parameters). 

\begin{figure}
  \centering
  \includegraphics[width=\columnwidth]{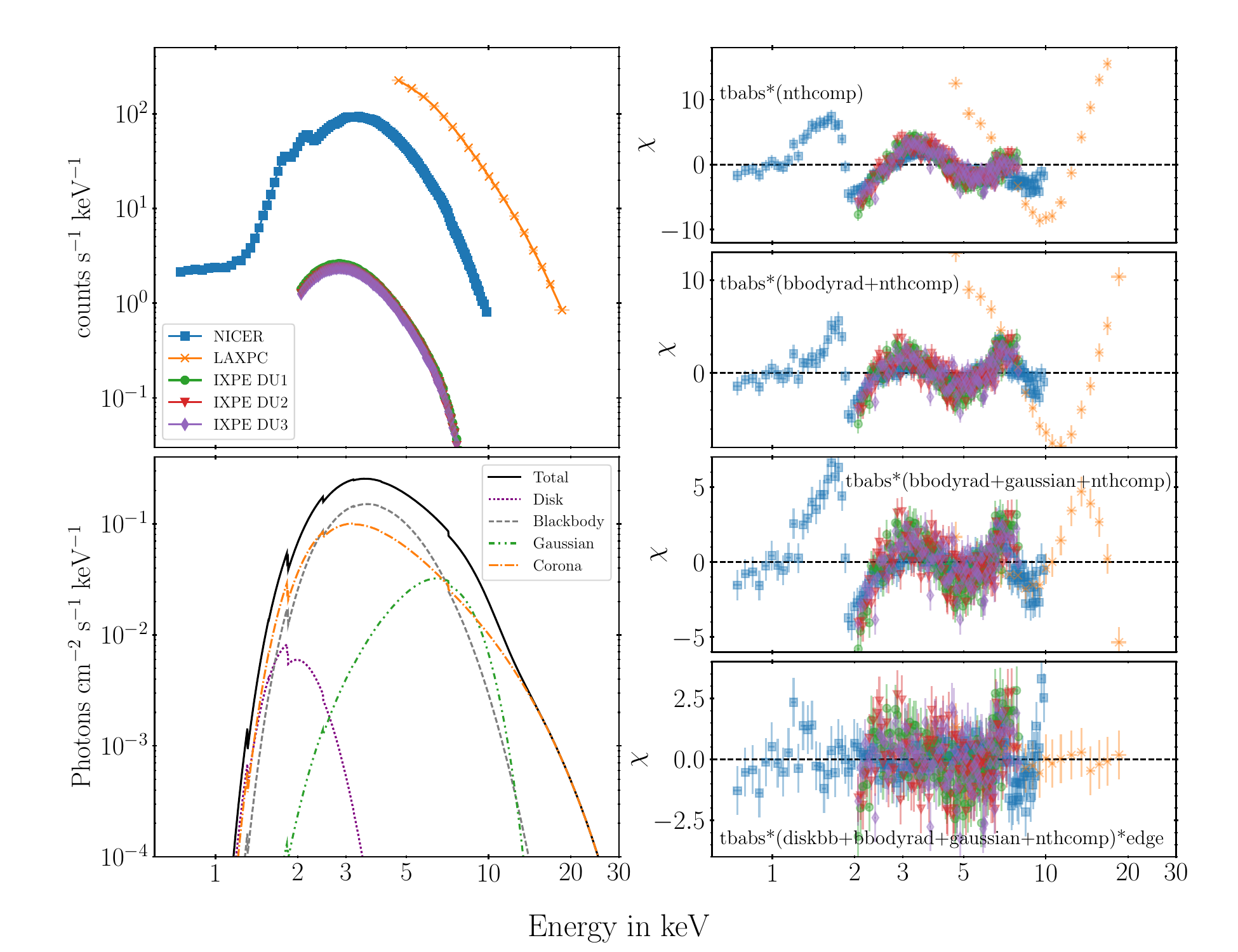}
  \caption{\textit{Top-left}: Count spectra of \src\ as observed during the NB by various instruments (depicted in different colors and markers indicated in the legend). \textit{Bottom-left}: Best fit model (\texttt{tbabs*(diskbb+bbodyrad+nthcomp+gaussian)*edge}) used to fit the count spectra is shown in the black solid curve, individual components are shown as various lines (indicated in the legend). \textit{Right panels}: Residuals ($\chi={\rm \frac{data-model}{error}}$) are shown for various intermediate models (noted in individual panels and described in Section~\ref{ssec:spec_mod}) and the residuals for the best fit model are shown in bottommost-right panel. }\label{fig:spec}
\end{figure}

Noting a high absorption column modeled in the literature of the source \citep[e.g.][]{Church2006A&A...460..233C, Iaria2006ChJAS...6a.257I}, we modeled the absorption column with \texttt{tbabs} and used the abundances from \cite{Wilms2000ApJ...542..914W} and cross-sections from \citet{Vern1996ApJ...465..487V}. Initially, we modeled spectra from individual observatories to identify the minimum number of additive components required to sufficiently model them.  
We fit the spectra with \texttt{tbabs*nthcomp} which assumes that the non-thermal component arises from a Comptonizing medium \citep{nthzdz1996MNRAS.283..193Z,nthzycki1999MNRAS.309..561Z}, which is up-scattering photons from an accretion disk. We observed that this combination resulted in an unacceptable fit statistic ($\chi^2$/degrees of freedom or d.o.f.=4233/601). The residuals clearly indicated the presence of a softer component, which we modeled with a blackbody (\texttt{bbodyrad}). The inclusion of a blackbody component resulted in a lower test statistic ($\chi^2$/d.o.f.=2531/599) but the fit was still not acceptable. The \nicer\ and \ixpe\ spectra indicated the presence of iron line emission that was modeled with a \texttt{gaussian}. For the \astr-LAXPC data, the normalization of the \texttt{gaussian} component was observed to be lower than other instruments (which could be due to calibration mismatch). 
The resulting $\chi^2$ of 1877 for 595 d.o.f. 
was still not acceptable, where the residuals indicated the presence of a softer component and possibly an edge at 1.8~keV.  We described this component as an accretion disk modeled with a soft multicolor blackbody (i.e., \texttt{diskbb}) and tied the inner disk temperature to the seed photon temperature for the \texttt{nthcomp} component ($\chi^2$/d.o.f.=698.6/594). 
After the inclusion of \texttt{diskbb}, the residuals at the highest energies were accounted for as now the Comptonized emission could describe the emission at higher energies without the need to compensate at softer energies\footnote{We note that the current spectral model is similar to the Western model with an addition of a soft and truncated accretion disk to explain the excess in the softer energies ($\lesssim2$~keV). This particular combination of models has also been tested in previous spectral investigations of \src\ by \citetalias{Bhargava2023ApJ...955..102B}, where a comparison to the Eastern model counterpart (i.e., softest thermal component is a blackbody while component $>1$~keV is the accretion disk) is discussed extensively.}. 
We note that the reduced $\chi^2$ is not acceptable at this stage and the fitting statistic of the \nicer\ spectra could be further reduced by modeling the edge-like residuals with a multiplicative component \texttt{edge}. 
After the inclusion of the edge component, the test statistic of $\chi^2$/d.o.f.=670.7/592 is acceptable (as compared to the previous case of model without the edge using either Akaike Information Criterion (AIC) or Bayesian Information Criterion (BIC)). 
The spectral parameters are noted in Table~\ref{tab:spec_pars}. To compute the confidence intervals for various spectral parameters, we sampled the parameter space using Markov Chain Monte Carlo sampler in \textsc{xspec}. We used the Goodman-Weare algorithm and initialized 30 walkers close to the best fit parameters and ran for a total 9000 steps after a 1980 burn-in steps. The confidence intervals were determined from the marginalized distribution of the sampled parameter space.  
We also show the count spectra from various instruments, the unfolded model and residuals from all the spectra at various stages of spectral modeling in Figure~\ref{fig:spec}. 

\begin{table}
  \centering
  \caption{Spectral modeling of \src\ during the NB. The model decomposition used for the following table is \texttt{constant*tbabs*(diskbb+bbodyrad+nthcomp+gaussian)*edge}. }\label{tab:spec_pars}
  \begin{tabular}{l|c|c|c}
    \hline
    Component & Parameter & Unit & Value \\ \hline
    \texttt{tbabs} & $n_{\rm H}$ & $10^{22}$ cm$^{-2}$ & $9.62_{-0.14}^{+0.09}$ \\ \hline
    \texttt{diskbb} & kT$_{\rm disk}$ & keV & $0.212_{-0.006}^{+0.003}$ \\
     & Norm$^\ast$ & $10^{5}$ & $9.9_{-1.5}^{+1.6}$ \\\hline
    \texttt{bbodyrad} & kT$_{\rm bb}$ & keV & $0.983_{-0.008}^{+0.008}$ \\
     &Norm$^\dagger$ & & $782_{-25}^{+18}$\\\hline
    \texttt{nthcomp} & $\Gamma$ &  & $2.25_{-0.03}^{+0.02}$ \\
     & kT$_{e}$ & keV& $3_{-1}^{+1}$ \\
     & kT$_{seed}$& keV& = kT$_{\rm disk}$ \\
     & Norm$^\ddagger$ & & $3.2_{-0.2}^{+0.1}$\\\hline
    \texttt{gaussian} & LineE & keV & $6.25_{-0.12}^{+0.06}$ \\
     & LineW & keV & $2.0_{-0.04}^{+0.07}$ \\
     & LineN$^\star$ & & $0.19_{-0.01}^{+0.02}$ $^a$ \\\hline
    \texttt{edge} & EdgeE & keV & $1.823_{-0.002}^{+0.005}$ \\
     & $\tau$&& $0.190_{-0.007}^{+0.011}$ \\\hline
    
    \texttt{constant} & C$_{DU1}$ &  & 0.648 \\
    &C$_{DU2}$&& 0.655\\
    &C$_{DU3}$&& 0.646\\
    &C$_{LAXPC}$&& 0.879\\\hline
    \multicolumn{4}{c}{Gain Parameters} \\\hline
    Instrument & Parameter & Unit & Value \\ \hline
    \ixpe-DU1& Slope & & 0.998 \\
    & Offset & eV & 2\\ \hline
    \ixpe-DU2& Slope & &  0.990\\
    & Offset & eV & 44\\ \hline
    \ixpe-DU3& Slope & & 0.996 \\
    & Offset & eV & 24\\ \hline
    \multicolumn{4}{c}{Fitting Statistic} \\\hline
    Instrument & \multicolumn{2}{c|}{Statistic} & Value (Stat/Bins) \\ \hline
    \nicer    &\multicolumn{2}{c|}{PG} & 275.8/146 \\
    \ixpe-DU1 &\multicolumn{2}{c|}{$\chi^2$} & 210.5/148\\
    \ixpe-DU2 &\multicolumn{2}{c|}{$\chi^2$} & 183.6/148\\
    \ixpe-DU3 &\multicolumn{2}{c|}{$\chi^2$} & 176.2/148\\
    LAXPC & \multicolumn{2}{c|}{$\chi^2$} &  6.34/20 \\ \hline
    \multicolumn{3}{c|}{Test Statistic: $\chi^2$/ d.o.f.} & 670.7/592 \\\hline
  \end{tabular}
  \begin{flushleft}
    Notes: \\
    $\ast$: The disk normalization is defined as Norm $=(R_{eff}/D_{10})^2*\cos \theta $, where $R_{eff}$ is the apparent inner radius of the accretion disk in km, $D_{10}$ is the distance to the source in units of 10~kpc, and $\theta$ is the inclination angle of the disk.  \\
    $\dagger $: The blackbody normalization is defined as $=(R/D_{10})^2$, where $R$ is the radius of the emission region in km.\\
    $\ddagger $: The normalization of \texttt{nthcomp} is defined as counts\,s$^{-1}$\,cm$^{-2}$\,keV$^{-1}$ at 1~keV. \\
    $\star$: The normalization of Gaussian component is the photons\,cm$^{-2}$\,s$^{-1}$ in the line. \\
    $a$: For LAXPC, the Gaussian normalization was estimated at $4_{-2}^{+8}\times10^{-3}$\,photons\,cm$^{-2}$\,s$^{-1}$. 
  \end{flushleft}
\end{table}

\subsection{Polarization in NB}\label{ssec:nb_pol}
Using the NB intervals as defined in Section~\ref{ssec:state_iden}, we investigated the polarization properties in a model independent manner (see Section~\ref{sssec:pcube}), and using the spectral decomposition from Section~\ref{ssec:spec_mod} we investigate the spectro-polarimetric properties of \src\ in Section~\ref{sssec:spec_pol}. 

\subsubsection{Model-independent analysis}\label{sssec:pcube}
\begin{figure}
  \centering
  \includegraphics[width=0.5\columnwidth]{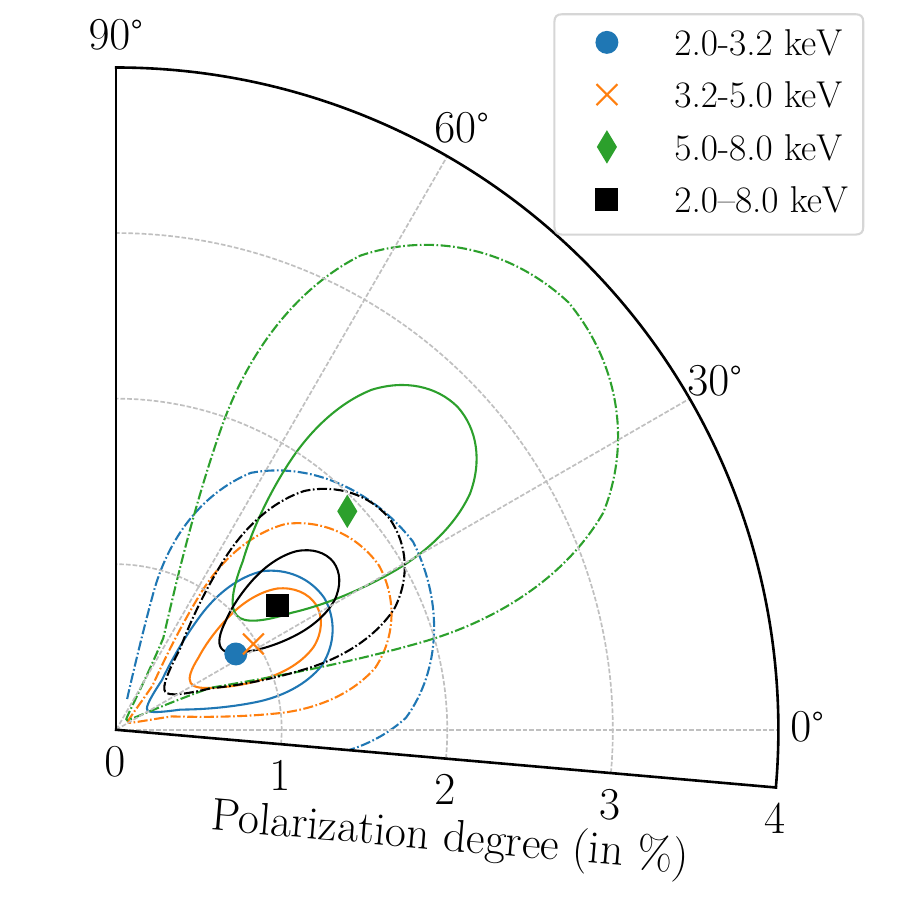}
  \caption{Model-independent polarization of \src\ during the NB. The markers depict the estimated PD and PA in the corresponding energy bins while the solid and the dashed lines correspond to 1 and 3$\sigma$ confidence intervals. Polarization was detected at $>3\sigma$ level in the 3.2--5~keV and 5--8~keV energy bands, and at $>4\sigma$ level for 2--8~keV, but not detected in the 2--3.2\,keV energy band. }\label{fig:pcube}
\end{figure}

To extract the model-independent polarization properties of \src\ in NB, we used the \textsc{pcube} algorithm to bin the \ixpe\ observation. Since the source was well above the background, we did not need to apply any background subtraction or rejection of events \citep[][]{dimarco2022AJ....164..103D}. We extracted the polarization information for individual detectors and then combined them  to enhance the measurement statistics.  We measured a PD of $1.3\pm0.3$\% and a PA of $38\pm6^\circ$ in the 2--8~keV band for the NB intervals. Since we have a highly significant detection, we divided the \ixpe\ operational band into three logarithmically spaced energy bins to extract the model-independent polarization properties. The full band and energy-dependent polarization properties are reported in Table~\ref{tab:pol_prop} and depicted in Figure~\ref{fig:pcube}. 
The polarization was not significant ($<3\sigma$) in the softest energy band (2.0--3.2~keV) while it is $>3\sigma$ in the 3.2--5.0~keV and 5.0--8.0~keV bands. 
We also estimated the polarization from the HB and FB/EFB intervals individually (grey and cyan intervals in Figure~\ref{fig:lc}, respectively) and found that the significance of the measurement was $<3\sigma$. Therefore, we placed upper limits on the X-ray polarization during these segments of the \ixpe\ observation for the full 2--8~keV band using \textsc{pcube}. We found  at  $3\sigma$ confidence, the HB PD is $<$1.8\% 
for a livetime of 42.46~ks and an FB PD is  $<$4.3\% for a livetime of 8.39~ks.

\begin{table}
  \centering
  \caption{Polarization of \src\ during the NB. 1$\sigma$ confidence intervals are reported. }\label{tab:pol_prop}
  \begin{tabular}{|l|cc|}
    \multicolumn{3}{c}{Model-independent polarization} \\\hline
    Energy band (keV) & PD (\%) & PA ($^\circ$) \\ \hline
    2.0--8.0 & $1.3\pm0.3$ & $38\pm6$ \\
    2.0--3.2 & $0.9\pm0.4$ & $32\pm14$\\
    3.2--5.0 & $0.9\pm0.3$ & $32\pm9$\\
    5.0--8.0 & $1.9\pm0.5$ & $43\pm9$\\ \hline

  \end{tabular}

\end{table}

\subsubsection{Spectro-polarimetric modeling}\label{sssec:spec_pol}
As shown in Figure~\ref{fig:spec}, the 2--8~keV emission has contribution from the blackbody, corona and the Gaussian emission components. To investigate the spectro-polarimetric properties of \src\ during the NB, we used the joint spectral fit from Section~\ref{ssec:spec_mod} and included the Stokes Q and U spectra from \ixpe\ to model the polarimetric properties.  We report 1$\sigma$ confidence intervals for the estimated polarization parameters. For this exercise, we fixed the spectral parameters (including the cross-normalization constants) and only allowed polarimetric parameters to be free. To test if the complete emission can be characterized by a single polarization component, we applied \texttt{polconst} and \texttt{pollin} to all additive components (the model description for constant polarization: \texttt{constant*tbabs*polconst*(diskbb+bbodyrad+nthcomp+gaussian)*edge}, and for linearly dependent polarization, \texttt{polconst} was replaced with \texttt{pollin}). 
In case of \texttt{polconst}, we note that the $\chi^2$ was 1563.8 for 1502 d.o.f. and the PD is $1.1\pm0.2\%$ for a PA of $33\pm5^\circ$. Since the polarization from the polarization cube does not indicate an energy dependence for the polarization angle, we kept the slope of the polarization angle fixed at zero for the \textsc{pollin} model. The $\chi^2$ for linear dependence of polarization on energy was 1560.5 for 1501 d.o.f. and the slope of the PD is $0.33_{-0.18}^{+0.08}$ which is consistent with zero within 2$\sigma$, indicating that while there may be a marginal dependence of the PD on the energy, our data are not entirely conclusive. 

The energy dependence of X-ray polarization as measured using \textsc{pcube} suggests that the softer energy bands have a marginal detection of polarization. To test if the observed model-independent polarization can be explained by a polarization from individual components, we applied \texttt{polconst} to \texttt{nthcomp}, \texttt{diskbb} and \texttt{bbodyrad}. Due to the marginal nature of soft X-ray polarization, we found that the fit is insensitive to any PD value of the disk component, therefore, we keep it fixed to 0.5\% which is the expected PD from an accretion disk \citep[e.g.][]{Li2009ApJ...691..847L} at an inclination angle of 35$\degr$ \citep{miller2016ApJ...822L..18M}. Additionally, we note that we obtain an acceptable fit if the PD of the blackbody component is fixed to zero and the PD and PA of the Comptonized emission is left free ($\chi^2/$dof = 1565.9/1502). In this case, we measured the PD of the Comptonized emission to be $2.76\pm0.5\%$, with a PA of $33\pm5^\circ$. An alternative test could be that the polarized emission could instead arise from the BL/SL (i.e. the component modeled with a blackbody). We obtained an equally acceptable fit ($\chi^2$=1566.0) in this case with a PD=$1.9\pm0.4$\% and PA=$32\pm5^\circ$.  

\section{Discussion} \label{sec:disc}

X-ray and radio observatories of \src\ taken on 2024 August 12--16 show that the source traced its complete Z-track, demonstrating rapid spectral state transitions \citep[][]{Jonker2000ApJ...537..374J, fridriksson2015}. \src\ traversed from the FB to the HB in less than a day (from MJD 60535.9 where it was in the FB to MJD 60536.2 where it entered the HB and stayed for about a day, see Figure~\ref{fig:lc}). 
However, for the majority of our observations, the source resided in the NB.

\subsection{Polarization in GX~340$+$0}

\src\ has been observed over two epochs with \ixpe: once in 2024 March where it was observed to be in the HB \citep[\citetalias{Bhargava2024arXiv240519324B},][]{LaMonaca2024arXiv241000972L} and again in 2024 August where it showed a complete tracing of the Z-track, but spending the majority of the time in the NB. In this work we have investigated the polarimetric properties of the source in the NB using the 2024 August observation (see \citetalias{Bhargava2024arXiv240519324B} for a detailed discussion of the 2024 March observations). The observed 2--8~keV PD during the NB ($1.3\pm0.3$\%, see Table~\ref{tab:pol_prop}) is lower than the PD measured during the HB \citepalias[4\%;][]{Bhargava2024arXiv240519324B}, but the PAs are consistent across both branches. The similarity in the PA indicates that the emission component in the hard X-rays is likely similar across both branches. Assuming the polarization is from a single component, i.e., a Comptonizing medium in both cases, would suggest that the emission has depolarized as the source moves from HB \citepalias[$9$\%;][]{Bhargava2024arXiv240519324B} to NB (2.4\%; Table~\ref{tab:pol_prop}). A similar decrease in net polarization was observed in simulations by \citet{gnarini2022MNRAS.514.2561G}, where the hard state polarization (similar to the HB in Z-sources) was observed to be higher than the soft state polarization (similar to the NB/FB/EFB in Z-sources).  
We also estimated the polarization properties for the HB and the FB/EFB segments in the present observation but the duration was not sufficient to detect polarization significantly, with a minimum detectable polarization at 99\% confidence (MDP99) of 1.7\% for HB and 3.8\% for FB/EFB respectively.  The MDP99 of the HB  was lower than the reported value in \citetalias{Bhargava2024arXiv240519324B} and \citet{LaMonaca2024arXiv241000972L} but we only obtain an upper limit of 1.8\% at a $3\sigma$ confidence interval. This could be due to either the inclusion of the hard apex in the current analysis or that the source spent insufficient time in the deepest part of the HB for the polarization to be detected significantly (which happened in the 2024 March \ixpe\ observation). 
The source was observed in FB/EFB for short duration (8.39~ks) which is not sufficient for a significant detection of polarization (as indicated by a large MDP99). Our observations place a $3\sigma$ upper limit of 4.3\% on the PD during these states. The FB/EFB observations of Cyg like Z-sources are limited to the observation analysed in this work while Sco X-1 shows a PD of $1.0\pm0.2$\% in its soft apex+FB \citep{monaca2024ApJ...960L..11L}. 

In the softer energies, the X-ray polarization was not significantly detected.  And therefore the slight discrepancy in the PA between 2-2.5~keV and 2.5--8 keV reported by \citetalias{Bhargava2024arXiv240519324B} and \citet{LaMonaca2024arXiv241000972L} during the HB was not constrained in the NB. According to \citetalias{Bhargava2024arXiv240519324B}, the accretion disk emission in 2--2.5~keV was the primary contributor towards the PA difference. In the spectral modeling of the NB data (Section~\ref{ssec:spec_mod}), we note that the disk appeared more truncated than during the HB \citepalias[e.g.][]{Bhargava2023ApJ...955..102B} with a lower contribution in the 2--8 keV band (the flux from \texttt{diskbb} component in computed using the parameters in \citetalias{Bhargava2023ApJ...955..102B} is 5 times higher than disk flux using parameters from Table~\ref{tab:spec_pars}) 
Therefore, a non-detection of the PA discrepancy is consistent with the spectro-polarimetric decomposition suggested by \citetalias{Bhargava2024arXiv240519324B}.

The X-ray polarization during the NB in \src\ is similar to the other Z-sources, e.g., Cyg X-2: $1.85\pm0.87\%$ \citep{Farinelli2023MNRAS.519.3681F}, XTE J1701$-$462: $<1.5\%$  \citep[at 99\% confidence;][]{Cocchi2023A&A...674L..10C}, GX 5$-$1: $2.0\pm0.3\%$ \citep{Fabiani2024A&A...684A.137F} and Sco X-1: $1.0\pm0.2\%$ \citep[][]{monaca2024ApJ...960L..11L}. 
Out of these sources, only XTE~J1701$-$462 and GX~5$-$1 have been observed in the HB as well \citep[which also match the \src\ HB polarization][\citetalias{Bhargava2024arXiv240519324B}]{Cocchi2023A&A...674L..10C, Fabiani2024A&A...684A.137F}. The similarity of polarization levels in these sources over different states, along with a similarity in timing signatures \citep[e.g.][]{hasinger1989A&A...225...79H}, hint at a similarity in the geometry of emission components. 

The simulations of weakly magnetized NS LMXBs \citep[][]{gnarini2022MNRAS.514.2561G} suggest that typical polarization values in HB and NB (which would correspond to hard and soft states in \citealt{gnarini2022MNRAS.514.2561G}) match with a slab like-corona. But the simulations also suggest that the PA variation is drastically different for both states. The difference in the PA is due to the assumption that the disk is polarized, and has a significant contribution in the 2--8~keV. If the thermal component is unpolarized (e.g. a blackbody component from the surface), then the PA dependence on energy will be caused by a single component (i.e. Comptonized emission).  In case of \src, we find a similar PA in HB and NB which would support this assumption and motivate simulations with different geometrical parameters to explain the observations.  We note that the spectral parameters used in the simulations by \citet{gnarini2022MNRAS.514.2561G} are significantly different from the parameter estimates from the spectral modeling. But at present, these are the best cases for comparison. 

Alternatively, the polarized emission can also arise from a broad SL. 
But the emission from SL is expected to have a strong variation of PA as a function of energy and the expected PD of the SL would be typically 1.5\% or lower \citep[][]{Bobrikova2024arXiv240916023B} and since our observations have shown constant PA as a function of energy (at least greater than 3~keV for both the HB and NB), and a higher PD in HB we can rule out SL as a possible origin for polarized emission. Another possibility is that the polarized emission is arising from the reflection component. \citet{LaMonaca2024arXiv241000972L} have explored this for the HB observation of \src\ and suggested that it is difficult to disentangle if the observed PD is arising from solely reflection component or solely Comptonized emission. 




\subsection{Radio emission from GX~340$+$0 and its correlation with the X-rays}
During the first GMRT observation on 2024 August 12, \src\ was in the NB. No radio emission was detected at 700 MHz, with a 3-sigma upper-limit of 1.1 mJy. During the second GMRT observation 2 days later (on 2024 August 14), \src\ had transitioned to the HB branch and the 1.26 GHz observations tentatively detected \src\ with a flux density of 4.5$\pm$0.7 mJy. At the start of the ATCA observations the next day (2024 August 15), the target was initially within the FB, transitioning back to the NB during the observation (Figure~\ref{fig:lc}). The ATCA observations detected radio emission from \src\ with a flux density of $0.70 \pm 0.05$\,mJy at 5.5\,GHz and $0.59 \pm 0.05$\,mJy at 9\,GHz. The measured flux densities could be consistent with emission from both a compact or transient jet, where the errors in the radio spectrum do not allow us to discriminate between the two. However, imaging the source on shorter timescales suggests that the radio emission was found to be fading at both frequencies during the first half of the ATCA observation (by a factor of 2--3) before remaining steady for the second half. Z-sources typically display fading radio emission as the source moves from the HB to the FB, with radio flares detected around the transition between the NB and HB, indicating the launching of ejecta \citep[e.g.,][]{2004Natur.427..222F,2019MNRAS.483.3686M}. Our results fit into this picture, where ejecta are expected to be launched over the HB$\rightarrow$NB transition, which then fade as the source moves towards the FB. We propose that the ATCA observations detected the fading of the flaring associated with this transition, before stabilizing as the source started to move back from the FB to the NB (when we expect it to begin to brighten; e.g., \citealt{2003ApJ...592..486B,2007ApJ...671..706M,2010A&A...512A...9B,2012A&A...546A..35C}). 

The results from the GMRT observations also fit the general picture of radio behavior of Z-sources; the non-detection during NB occurs due to a spectral turnover of the synchrotron emission at $>$700\,MHz, likely arising from free-free or synchrotron self-absorption \citepalias[see discussions in][]{Bhargava2024arXiv240519324B}. Z-sources are at their radio brightest during the HB, where the emission likely arises from a steady jet \citep[e.g.,][]{2007ApJ...671..706M}. As such, the tentative detection at 1.26\,GHz on 2024 August 15 as the source moves along the HB could be radio emission from a bright, steady jet.

\section{Conclusions} \label{sec:conc}

Based on an extensive campaign of \src\ in X-ray and radio wavelengths during its evolution along its Z-track, we constrained the spectral and polarimetric properties of the source within the NB. Here we summarize some of the key properties of the source we have inferred from the analysis:

\begin{itemize}
    \item We estimated the broadband polarization (2--8~keV) of \src\ during the NB at $1.3\pm0.3\%$ with a PA of $38\pm6^\circ$.
    \item The spectral analysis of the source using various instruments suggested the presence of three continuum components in the spectra: a low temperature accretion disk, a hot blackbody component (perhaps from the surface), and a Comptonized emission, as well as a Gaussian emission component corresponding to reprocessed emission from the accretion disk.
    \item Spectro-polarimetric study of the observations suggests that the PD of the Comptonizing medium in NB is weaker as compared to HB. 
    \item The X-ray polarization was significantly detected at energies $>3.2$~keV and the PA of the source is similar in NB and HB, suggesting a similar origin of the polarized emission in the source in both branches, perhaps from the Comptonizing medium. The additional component whose presence is hinted in HB is apparently not emitting strongly in the NB. 

\end{itemize}
\begin{acknowledgments}
  This research used data products provided by the \textit{IXPE} Team (MSFC, SSDC, INAF, and INFN). 
  This work makes use of data from the AstroSat mission of the Indian Space Research Organisation (ISRO), archived at the Indian Space Science Data Centre (ISSDC). The article has used data from the SXT and the LAXPC developed at TIFR, Mumbai, and the AstroSat POCs at TIFR are thanked for verifying and releasing the data via the ISSDC data archive and providing the necessary software tools. 
  This work was supported by NASA through the \nicer\ mission and  the Astrophysics Explorers Program. 
  This work has made use of the data from the \hxmt\ mission, a project funded by the China National Space Administration (CNSA) and the Chinese Academy of Sciences (CAS).

  This research has also made use of data and/or software provided by the High Energy Astrophysics Science Archive Research Center (HEASARC), which is a service of the Astrophysics Science Division at     NASA/GSFC and the High Energy Astrophysics Division of the Smithsonian Astrophysical Observatory. We thank the staff of the GMRT that made these observations possible. GMRT is run by the National Centre for Radio Astrophysics of the Tata Institute of Fundamental Research. The Australia Telescope Compact Array is part of the Australia Telescope National Facility (\url{https://ror.org/05qajvd42}), which is funded by the Australian Government for operation as a National Facility managed by CSIRO. We acknowledge the Gomeroi people as the Traditional Owners of the ATCA observatory site. We thank Jamie Stevens and the ATCA staff for making the observations possible. YB and AB would like to especially thank Dr Shriharsh Tendulkar for valuable discussion on radio observations and insights into the emission from the neutron stars. 
 TDR is a IAF research fellow. MN is a Fonds de Recherche du Quebec – Nature et Technologies (FRQNT) postdoctoral fellow.
  LZ acknowledges support from the National Natural Science Foundation of China (NSFC) under grant 12203052. JH acknowledges support for this work from the IXPE Guest Investigator program under NASA grant 80NSSC24K1746.
\end{acknowledgments}

%

\vspace{5mm}
\facilities{\ixpe, \astr, \nicer, \hxmt, \gmrt, \atca}


\software{\texttt{astropy} \citep{2018AJ....156..123A},  
        \texttt{ixpeobssim} \citep{ixpeobssim2022SoftX..1901194B},
        \texttt{CASA} \citep{CASA2022PASP..134k4501C},
        \texttt{Stingray} \citep{2019JOSS....4.1393H,2019ApJ...881...39H}
         }



\appendix

\section{Effect of softer color definition on the HID}\label{appsec:soft_HID}

\begin{figure}
  \centering
    \includegraphics[width=0.5\columnwidth]{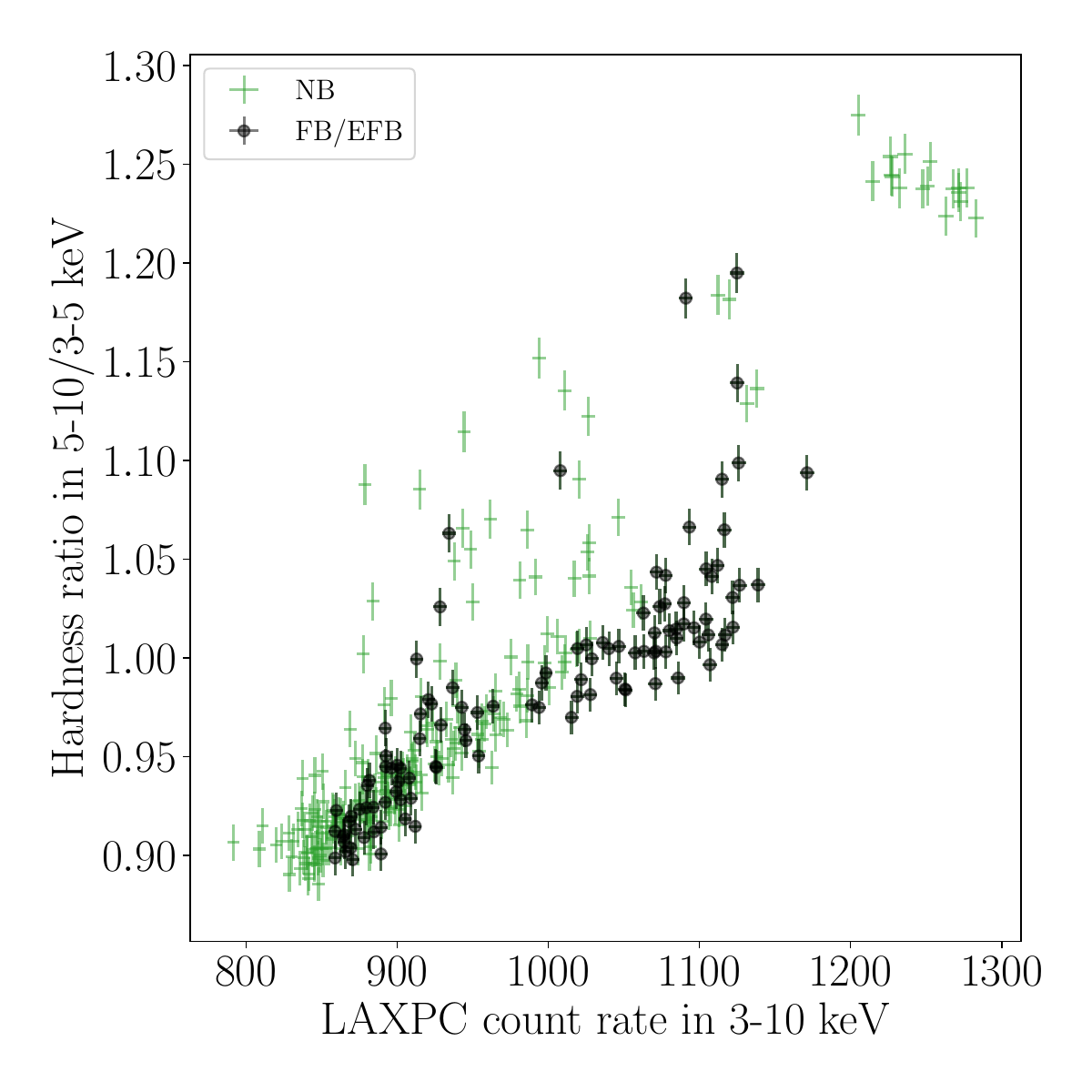}
    \caption{HID of \src\ as seen by \astr-LAXPC in the soft energy range. To highlight the overlap of the FB/EFB and the NB in the soft energy range, we depict the FB/EFB (as seen in Figure~\ref{fig:hid}) as black circles while the green points correspond to the NB and hard apex. A definition using a soft X-rays of the hardness ratio is unable to differentiate between the NB and the FB indicating that the distinctive shape of the FB/EFB is mainly due to changes in the 10--20~keV count rate. }\label{fig:hid_astrosat_soft}
  \end{figure}

We note that the FB and the EFB are not identifiable in the HID created using the soft X-ray bands (i.e., using the hardness ratio from any subset of energy bands in 3--10~keV, see figure~\ref{fig:hid}). We construct the HID from the \astr-LAXPC observation in the energy bands used for \nicer, i.e., 3--10~keV for the count rate and HR between the 5--10 and 3--5~keV bands (shown in figure~\ref{fig:hid_astrosat_soft}). We observe that a single track similar to the NB is visible in the HID. The points corresponding to the FB and the EFB are identified from Figure~\ref{fig:hid} and marked as black circles in Figure~\ref{fig:hid_astrosat_soft}. Additionally, the \astr\ observation shows a significant overlap with \nicer\ observation 
during one of the FB excursions.  The light curves from both observations seem to indicate the presence of a flare (during the FB) but the HR between the 5--10 and 3--5~keV bands was similar for both the FB and NB. Therefore it is challenging to accurately identify any FB intervals solely based on data from the soft X-ray instruments used here (i.e., \nicer\ or \ixpe).

\bibliography{ref}{}
\bibliographystyle{aasjournal}

\end{document}